\documentclass{aa}
\usepackage{txfonts}
\usepackage{graphicx}
\def\S{\mbox{S255~NIRS\,3}}

\def\Msun{\mbox{$M_\odot$}}
\def\Lsun{\mbox{$L_\odot$}}

\def\HII{H{\sc ii}}

\def\kms{\mbox{km~s$^{-1}$}}
\def\cmc{cm$^{-3}$}

\def\Log{\mbox{\rm Log$_{10}$}}

\def\e{{\rm e}}

\def\tho{\mbox{$\theta_0$}}
\def\ymax{\mbox{$y_{\rm max}$}}

\begin{document}

\title{
 Radio outburst from a massive (proto)star
\thanks{Based on observations carried out with the VLA, IRAM/NOEMA, and ALMA.}
}
\subtitle{When accretion turns into ejection}
\author{
        R.~Cesaroni\inst{1}
        \and
        L.~Moscadelli\inst{1}
        \and
        R.~Neri\inst{2}
        \and
        A. Sanna\inst{3}
        \and
        A. Caratti o Garatti\inst{4}
        \and
        J. Eisl\"offel\inst{5}
        \and
        B. Stecklum\inst{5}
        \and
        T. Ray\inst{4}
        \and
        C.~M.~Walmsley\thanks{This article is dedicated to the memory of Malcolm Walmsley, who
passed away before the present study could be completed. Without his insights and
enlightened advice this work would have been impossible. We will always
remember all the stimulating discussions with him, as well as his delightful
personality.}
}
\institute{
 INAF, Osservatorio Astrofisico di Arcetri, Largo E. Fermi 5, I-50125 Firenze, Italy
           \email{cesa@arcetri.astro.it}
\and
 Institut de Radioastronomie Millim\'etrique (IRAM), 300 rue de la Piscine, F-38406 Saint Martin d'H\`eres, France
\and
 Max Planck Institut f\"ur Radioastronomie, Auf dem H\"ugel 69, D-53121 Bonn, Germany
\and
 Dublin Institute for Advanced Studies, Astronomy \& Astrophysics Section, 31 Fitzwilliam Place, Dublin 2, Ireland
\and
 Th\"uringer Landessternwarte Tautenburg, Sternwarte 5, D-07778 Tautenburg, Germany
}
\offprints{R. Cesaroni, \email{cesa@arcetri.astro.it}}
\date{Received date; accepted date}

\abstract{
Recent observations of the massive young stellar object \S\
have revealed a large increase in both methanol maser flux density
and IR emission, which have been interpreted as the result of an accretion
outburst, possibly due to instabilities in a circumstellar disk. This indicates
that this type of accretion event could be common in young/forming early-type stars
and  in their lower mass siblings, and supports the idea that accretion
onto the star may occur in a non-continuous way.
}{
As accretion and ejection are believed to be tightly associated phenomena, we
wanted to confirm the accretion interpretation of the outburst in \S\ by detecting
the corresponding burst of the associated thermal jet.
}{
We monitored the radio continuum emission from \S\ at four bands using
the Karl G. Jansky Very Large Array.  The millimetre continuum emission
was also observed with both the Northern Extended Millimeter Array of IRAM
and the Atacama Large Millimeter/submillimeter Array.
}{
We have detected an exponential increase in the radio flux density from 6
to 45~GHz starting right after July 10, 2016, namely $\sim$13~months after
the estimated onset of the IR outburst. This is the first ever detection of
a radio burst associated with an IR accretion outburst from a young stellar object.
The flux density at all observed  centimetre bands can be reproduced with a
simple expanding jet model.  At millimetre wavelengths we infer a marginal
flux increase with respect to the literature values and we show this is due
to free-free emission from the radio jet.
}{
Our model fits indicate a significant increase in the jet opening angle
and ionized mass loss rate with time. For the first time, we can estimate
the ionization fraction in the jet and conclude that this must be low
($<$14\%), lending strong support to the idea that the neutral component
is dominant in thermal jets. Our findings strongly suggest that recurrent
accretion+ejection episodes may be the main route to the formation of
massive stars.
}

\maketitle

\section{Introduction}
\label{sint}

The accretion phenomenon has recently gained importance not only in
the formation process of solar-type stars, but also across the whole stellar
mass spectrum. Growing evidence of the presence of disks around B-type
and even O-type stars lends strong support to the hypothesis that even
the most massive stars may form through disk-mediated accretion, as
the latter can overcome their powerful radiation pressure (Krumholz et
al.~\cite{krum07a}; Kuiper et al.~\cite{kuip10}; Peters et al.~\cite{pet10}).
While most theoretical models assume a smooth, steady accretion flow onto
the star, it is very likely that in real life the flow is much more
irregular, due to disk inhomogeneities and fragmentation.
This scenario is suggested by the detection of sudden increases
in  the stellar luminosity in low-mass (proto)stars, which hint at episodic
accretion events. These phenomena are known as FUor and EXor outbursts
and consist of an increase of up to 6~mag in the young stellar object (YSO)
luminosity in the optical and near-IR, which corresponds to an increase in
the mass accretion rate of up to $\sim$$10^{-4}~M_\odot$\,yr$^{-1}$
(Audard et al.~\cite{aud}).  Correspondingly, these accretion events are
believed to trigger an increase in the mass ejection rate.

While recent studies have corroborated the idea that accretion bursts
take place through a broad range of stellar masses (Acosta-Pulido et
al.~\cite{acos}; Caratti~o~Garatti et al.~\cite{caga11}; Contreras
Pe\~na et al.~\cite{cope}), it is not obvious that they should also occur
 in early-type stars, namely for stellar masses in excess of
$\sim$6~\Msun. However, evidence of such a burst was reported by Tapia et
al.~\cite{tapia} in V723~Car, a YSO with a relatively high luminosity
($\sim$$4\times10^3$~\Lsun).

More recently, Fujisawa et al.~(\cite{fuji}) detected a powerful methanol
maser flare at 6~GHz towards the 20~\Msun\ YSO \S, located at a
distance of 1.78$^{+0.12}_{-0.11}$~kpc (Burns et al.~\cite{burns}). Since
CH$_3$OH~class~II masers are believed to be radiatively pumped, follow-up
observations at IR wavelengths were performed, which indeed revealed a
dramatic brightening of the source of 2.9~mag and 3.5~mag at
K and H bands, respectively (Stecklum et al.~\cite{steck}), and a corresponding increase
in bolometric luminosity of $\sim$$1.3\times10^5~L_\odot$ (Caratti~o~Garatti
et al.~\cite{cagana}). Our study has shown that such an increase is likely
driven by an accretion burst of $\sim$$5\times10^{-3}~M_\odot$~yr$^{-1}$ that occurred around mid-June 2015. This result is consistent with the proposed
existence of a disk-jet system in \S\ (Wang et al.~\cite{wang}; Boley et
al.~\cite{bole}; Zinchenko et al.~\cite{zin15}), which may be mediating the
accretion process.

Finally, our high angular resolution observations (from 1~mas to 1\arcsec)
of the methanol maser flare revealed a substantial
change in the circumstellar environment of the bursting source. In
particular, the main pre-burst maser cluster, located close to the radio continuum peak,
is no longer detected during the burst. Instead, a new extended plateau has
formed at a projected separation of 500--1000~au from the YSO (Moscadelli
et al.~\cite{mosca17}).

Since November 2015 we have been performing multi-epoch observations of the
source in a variety of tracers from the IR to the radio, with a number of
instruments such as the VLT, IRTF, 2.2\,m Calar Alto, Effelsberg, SOFIA, and
the GMRT. In particular, we are monitoring the centimetre and millimetre
continuum emission by means of the Karl G. Jansky Very Large Array (VLA), the
Northern Extended Millimeter Array (NOEMA) of IRAM, and the Atacama Large
Millimeter/submillimeter Array (ALMA). Our main goal was to establish
whether  the accretion burst had boosted the radio thermal jet visible in
archival VLA images of \S\ (see Fig.~\ref{fbeaf}a). Since accretion and
ejection are believed to be closely related phenomena during the star
formation phase, our expectation was to detect a sudden brightening of
the radio emission from the jet itself as a consequence of the accretion
burst revealed from the IR brightening. Indeed, {our monitoring resulted
in the first ever detection  of a radio burst associated with an IR outburst},
and we report on our finding in the present article.

In Sect.~\ref{sobs} we give the observational details,
while the results obtained are described in Sect.~\ref{sres} and analysed
through a suitable model in Sect.~\ref{sdis}. Finally, the conclusions
are drawn in Sect.~\ref{scon}.

\section{Observations and data analysis}
\label{sobs}

In the following, we report on the VLA data obtained in 2016 and the
ALMA and IRAM/NOEMA data acquired between December~2016 and February~2017.

In all cases, the phase centre  of the observations was set at the
coordinates $\alpha$(J2000)=$06^{\rm h}\,12^{\rm m}\,54\fs02$,
$\delta$(J2000)=17\degr\,59\arcmin\,23\farcs1.

\subsection{Very Large Array}

\S\ was observed using the VLA of the National Radio Astronomy
Observatory (NRAO) at five epochs during 2016 (project code: 16A-424): in
C~configuration on March 11, in B~configuration on July 10 and August 1,
and in A~configuration on October 15, November 24, and December 27.

Each observing epoch lasted 1.73~hours and recorded the signal at 6.0, 10.0,
22.2, and 45.5~GHz. We employed the capabilities of the WIDAR correlator,
which permits recording dual polarization across a total bandwidth per
polarization of 2~GHz, by using two IFs, each comprising eight adjacent
128~MHz sub-bands. The total observing bandwidth (per polarization) was
4.1~GHz (with 2 IFs) at 6.0 and 10.0~GHz, and 8.0~GHz (with 4 IFs) at 22
and 45~GHz.  The primary flux calibrator was 3C48, and the phase-calibrator
was J0559$+$2353, separated by $\sim$7\degr\ from the target.

Calibration and imaging were performed with the CASA\footnote{The
Common Astronomy Software Applications software can be downloaded at
http://casa.nrao.edu} software, by using different versions (from 4.3.1
to 4.5.2) of the JVLA data reduction pipeline. We estimate a calibration
accuracy of $\sim$10\%. The continuum images were constructed using natural
or
Briggs robust=0
weighting. Table~\ref{tmon} reports the mean synthesized
beam widths at half power, at the five observing epochs.  The rms noise
level of the continuum images is close to the expected thermal noise,
in the range 10--20~$\mu$Jy~beam$^{-1}$ at 6, 10~GHz, and 22~GHz, while
it is significantly higher, 50--100~$\mu$Jy~beam$^{-1}$ at 45~GHz, where
atmospheric phase noise is dominant. At 22~and~45~GHz, we self-calibrated
the compact continuum emission, improving the dynamic range of the image
by a factor of a few.

\subsection{IRAM/NOEMA}

We observed \S\ with the interferometer on January~23 and February~15,
2017. The array consisted of eight elements  in the compact D
configuration. The bandpass and phase calibrator were J0507+179, while flux
calibration was performed by means of observations of LkHa101. We used
the wideband correlator (WideX) to obtain a measurement of the continuum
emission at both 3.4 and 2~mm. For this purpose the receivers were tuned
to cover a 3.6~GHz wide band centred at 87 and 150~GHz. All observations
were performed with on-source integration times of 10~min. Observations
were carried out first at 3.4~mm, and were  immediately followed  by
observations at 2~mm.

Data calibration and imaging were made through standard procedures,
using the GILDAS package\footnote{The GILDAS software has been
developed at IRAM and Observatoire de Grenoble: 
http://www.iram.fr/IRAMFR/GILDAS}. All maps were made with natural
weighting, resulting in a synthesized beam with full widths at half
power of 5\farcs0$\times$3\farcs5 with PA=195\degr\ at 3.4~mm and
3\farcs0$\times$1\farcs9 with PA=0\degr\ at 2~mm in January, and
5\farcs1$\times$3\farcs3 with PA 193\degr\ at 3.4~mm and
2\farcs9$\times$1\farcs8 with PA=11\degr\ at 2~mm in February. The
continuum maps were obtained by visually inspecting the data cubes and
selecting channels with no evidence of line emission. These channels
were then averaged in the $u,v$ domain and pure continuum images were
made from the averaged $u,v$ data. The resulting continuum bandwidth
was 682~MHz at 3.4~mm and 379~MHz at 2~mm. The thermal noise level is
typically 0.1~mJy/beam at 3.4~mm and 0.3~mJy/beam at 2~mm. Flux
calibration is accurate to within 20\% at both wavelengths.

\subsection{Atacama Large Millimeter/submillimeter Array}

\S\ was observed with ALMA at band~3 on December 16, 2016.  The array
consisted of 45 antennas arranged in a configuration with baselines from
$\sim$20 to $\sim$400~m, which provides sensitivity to structures up to
$\sim$30\arcsec. Since our main goal was to image the continuum emission,
we configured the correlator to cover a total bandwidth of $\sim$8~GHz
with four units of 2~GHz, in double polarization, centred at 85.2, 87.2,
97.2, and 99.2~GHz. We chose the maximum possible spectral resolution
of 0.49~MHz in order to identify strong lines that could significantly
contaminate the continuum measurement.  For flux and bandpass calibration
we used the strong calibrators J0510$+$1800 and  J0750$+$1231, while the
phase calibrator  J0613$+$1708 was separated by less than
1\degr\ from our target. Calibration and imaging were performed with the
CASA software by employing the ALMA data reduction pipeline. We estimate
a calibration error of $\sim$20\%.

A data cube was created from each 2~GHz correlator unit with natural
weighting. In order to obtain a pure continuum image, we adopted the STATCONT
method by S\'anchez-Monge et al.~(\cite{statcont}), which automatically
identifies line-free channels in the cubes. In this way four continuum maps
were obtained at the four centre frequencies of the correlator units. The
synthesized half-power beam widths and the 1$\sigma$ RMS noise ranged
respectively  from 2\farcs21$\times$2\farcs12 to 2\farcs61$\times$2\farcs45,
and from 0.34 to 0.61~mJy/beam, depending on the frequency.

Finally, for each continuum image the flux density of \S\ was computed by
integrating the emission over the corresponding core (indicated with SMA1
by Zinchenko et al.~\cite{zin12}). Since pairs of correlator units are
close in frequency, we computed a mean value of the flux for each pair,
thus obtaining a continuum flux density of $\sim$73~mJy at 86.2~GHz and
$\sim$90~mJy at 98.2~GHz.

\section{Results}
\label{sres}

\subsection{Centimetre emission}
\label{scm}

\begin{figure}[h!]
\centering
\resizebox{8.5cm}{!}{\includegraphics[angle=0]{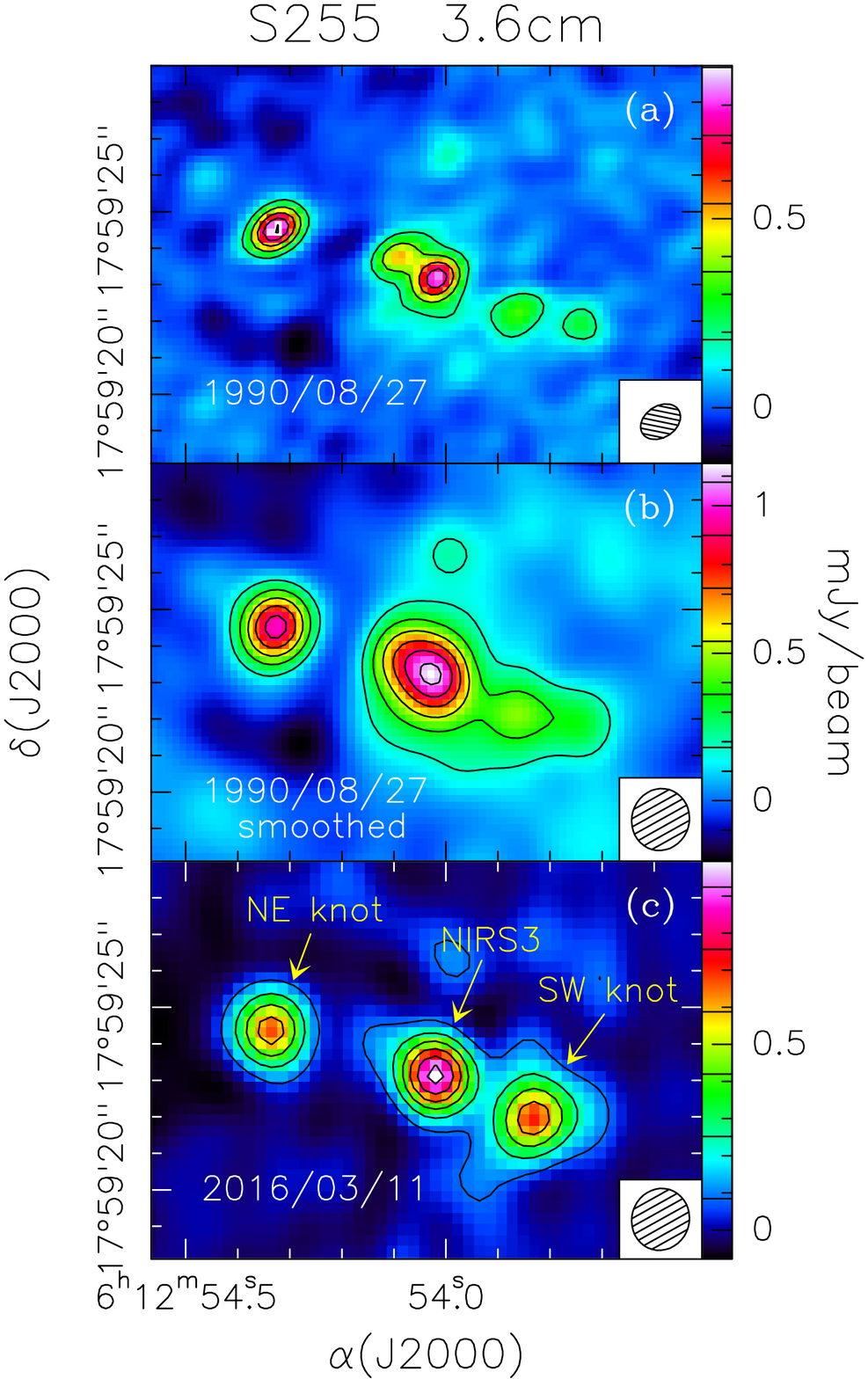}}
\caption{
{\bf a.} VLA archival image obtained on August 27, 1990, at 3.6~cm with the
C-array configuration.
The values of the contour levels are drawn in the corresponding colour scales
to the right. The full width at half power of the synthesized beam is shown
in the bottom right.
{\bf b.} Same as the top panel, after smoothing to the resolution of our new
3~cm image.
{\bf c.} Same as the other panels, but for our new 3~cm image obtained on
March 11, 2016, with the C-array configuration.  The bipolar structure
of the jet is shown, consisting of a central core associated with NIRS\,3 and two
main knots to the NE and SW.
}
\label{fbeaf}
\end{figure}

Our VLA monitoring started on March 11, 2016,  about 9~months after
the estimated onset of the outburst (mid-June 2015; see Caratti o Garatti
et al.~\cite{cagana}). As illustrated in Fig.~\ref{fbeaf}, comparison
with VLA archival data obtained at the same frequency on August 27, 1990
(project AM301), with similar angular resolution reveals some structural
change. In the middle panel of the figure we also show the archival map
after smoothing to the same angular resolution as our new data.

All images clearly outline an elongated structure consisting
of a central bright core, coincident with the NIRS\,3 infrared source,
plus two symmetrically displaced knots to the NE and SW. We interpret this
structure as a bipolar radio jet, identified by us for the first time.
Indeed, the jet lobes are oriented in the same direction as the axis of
the bipolar outflow identified by Zinchenko et al.~(\cite{zin15}).
Moreover, the two knots coincide with Fe{\sc ii} 1.64~$\mu$m emission
detected by Wang et al.~(\cite{wang}), which lends further support to
our interpretation.

The archival images in Figs.~\ref{fbeaf}a and \ref{fbeaf}b differ 
from that in Fig.~\ref{fbeaf}c in the SW lobe, which appears split into
two knots. Moreover, the central core looks fainter in our new image (taken on March 11, 2016) with
a total flux density $\sim$1.7 times less than that measured in 1990. A decrease
of the same amount is observed in the NE knot, whereas the total flux from
all the SW knots ($\sim$0.9~mJy) has remained basically the same.

We conclude that in 26~years the radio jet has undergone significant
morphological changes over an angular scale comparable to the size of
the SW lobe, i.e. $\sim$2\arcsec\ or $\sim$3600~au on the plane of the
sky. This implies that the ionized gas in the jet is moving at an average
speed on the plane of the sky of at least $\sim$660~\kms. However, the observed changes do not
support the occurrence of a radio burst at the current epoch because the radio emission from
the whole jet and, in particular, from the central core has not increased.
Unlike the SW lobe, the shape of the NE lobe has remained basically
the same, as has its distance from the central source. This suggests
that any ejection episode might be highly asymmetric, as also discussed 
by other authors (e.g. Melnikov et al.~\cite{mel08,mel09}; Caratti o
Garatti et al.~\cite{caga13}; Cesaroni et al.~\cite{cesa13}).

The second epoch of our VLA monitoring (July 10, 2016) has confirmed the
lack of brightening of the emission. The flux density from the
central core (and that of the other two knots) is unchanged, within
the uncertainties, with respect to that measured at the same frequency
in March. This can be seen in Table~\ref{tmon} where the fluxes measured
from NIRS\,3 during our monitoring are given.  The only significant change
is seen at 46~GHz, where the flux density is less intense in July than
in March. However, this decrease is likely due to the more extended
configuration (B-array) of the second-epoch observations, which resolves out
the dust emission from the hot molecular core enshrouding NIRS\,3.

\begin{table*}
\centering
\caption[]{
Flux densities measured towards the central core of \S\ with the VLA, IRAM/NOEMA, and ALMA at
different epochs, and corresponding mean synthesized beams. Months from March
to December refer to  2016, while January and February refer to 2017.
}
\label{tmon}
\begin{tabular}{ccccccccc}
\hline
\hline
date & Mar & Jul & Aug & Oct & Nov & Dec & Jan & Feb \\
\hline
$\nu$ & \multicolumn{8}{c}{$S_\nu$, HPBW} \\
(GHz) & (mJy, arcsec) & (mJy, arcsec) & (mJy, arcsec) & (mJy, arcsec) & (mJy, arcsec) & (mJy, arcsec) & (mJy, arcsec) & (mJy, arcsec) \\
\hline
6 & 0.87, 2.54 & 0.79, 0.998 & 1.1, 1.33 & 2.6, 0.358 & 4.8, 0.402 & 7.9, 0.352 & -- & -- \\
10 & 1.0, 1.64 & 1.2, 0.609 & 1.8, 0.765 & 3.6, 0.168 & 6.7, 0.235 & 11, 0.209 & -- & -- \\
22.2 & 2.7, 0.73 & 2.2, 0.354 & 3.1, 0.351 & 6.6, 0.0695 & 11, 0.108 & 18, 0.0969 & -- & -- \\
45.5 & 4.4, 0.412 & 1.9, 0.190 & 3.8, 0.195 & 14, 0.0437 & 21, 0.0455 & 27, 0.0800 & -- & -- \\
 86.2 & -- & -- & -- & -- & -- & 73, 2.5 & -- & -- \\
87 & -- & -- & -- & -- & -- & -- & 57, 4.2 & 64, 4.1 \\
 98.2 & -- & -- & -- & -- & -- & 90, 2.2 & -- & -- \\
150 & -- & -- & -- & -- & -- & -- & 175, 2.4 & 168, 2.3 \\
\hline
\hline
\end{tabular}
\end{table*}

\begin{figure*}
\centering
\resizebox{17.5cm}{!}{\includegraphics[angle=0]{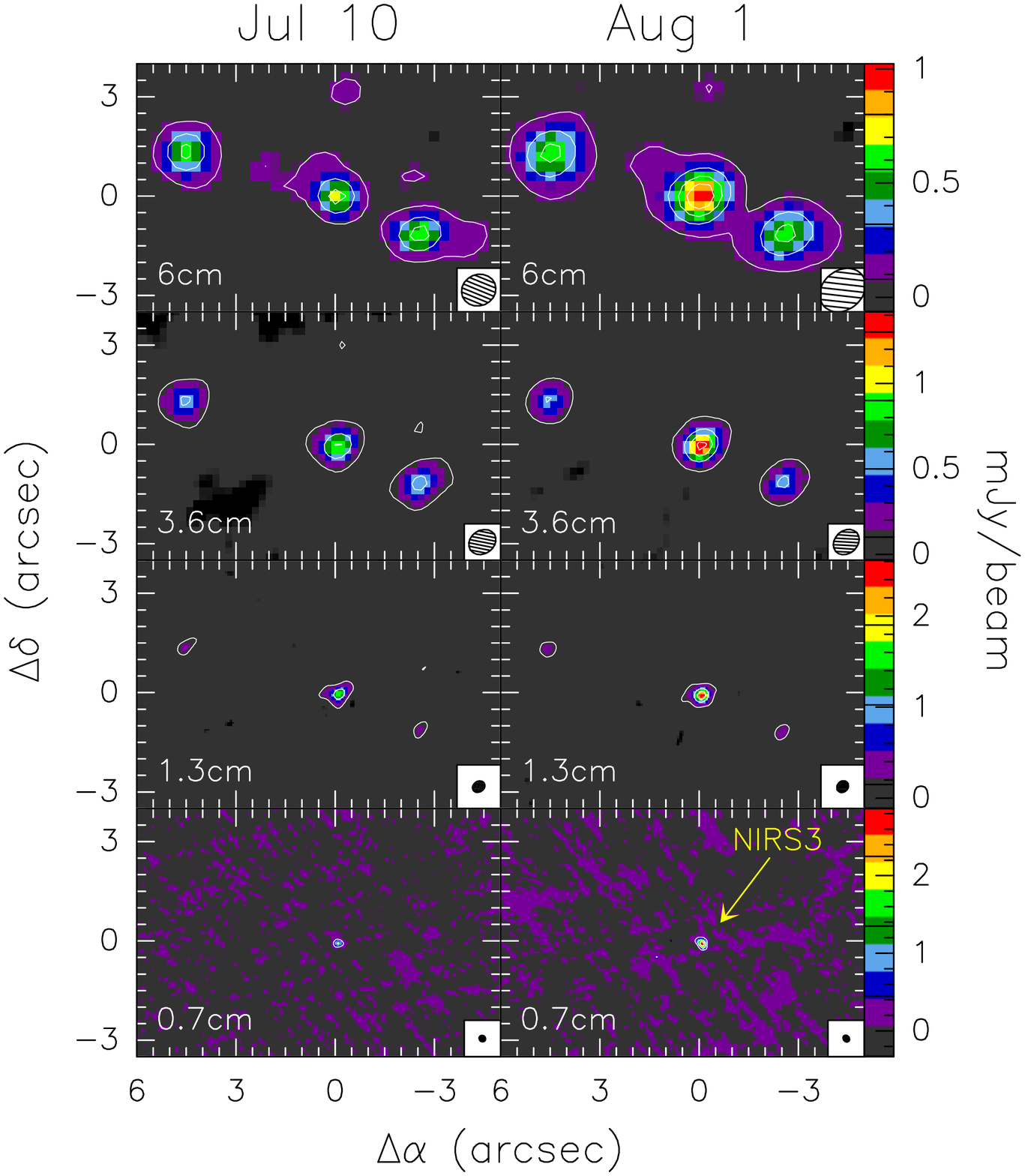}}
\caption{
Maps of the \S\ region at centimetre wavelengths obtained with the VLA on
July~10 (left panels) and August~1 (right panels), 2016. The offsets are
computed with respect to the phase centre of the observations, i.e.
$06^{\rm h}\,12^{\rm m}\,54\fs02$, 17\degr\,59\arcmin\,23\farcs1. The corresponding
synthesized beams are shown at the bottom right of each panel. The values
of the contour levels are marked in the colour scales to the right, and
in all cases the lowest contour is $>4\sigma$. We note the brightening of
NIRS\,3 at all frequencies.
}
\label{fjuau}
\end{figure*}

At the third epoch (August 1, 2016) we suddenly detected an increase in the
radio emission from the central core of NIRS\,3 at all observed frequencies. We note that at the
same epoch no changes were detected from the NE and SW knots. This brightening
is undoubtedly real and not due to a calibration error  because we observe
an increase of a factor $\sim$1.6 in the ratio between the flux of the central
core and that of the two radio knots. This can be easily seen by visual
comparison between the VLA maps made in July and those made in August
(see Fig.~\ref{fjuau}). We stress that the same array configuration (B)
was used at both epochs, hence the difference in flux cannot be attributed
to different u-v coverage.

Since no variation is detected from the NE and SW knots,
hereafter we will focus only on the central radio core, NIRS\,3, and we
 refer to it simply as ``the jet''.

It is interesting to note that the jet flux measured in August at 10~GHz
(1.8~mJy; see Table~\ref{tmon}) is equal to the value obtained from the
archival data in 1990 (Fig.~\ref{fbeaf}a). This might indicate
that the source underwent other outburst episodes in the past,
consistent with a scenario where accretion occurs mainly in an irregular way.
A similar conclusion was attained by Burns et al.~(\cite{burns}), who identify
the signposts of three distinct ejection events from \S.

\begin{figure}
\centering
\resizebox{8.5cm}{!}{\includegraphics[angle=0]{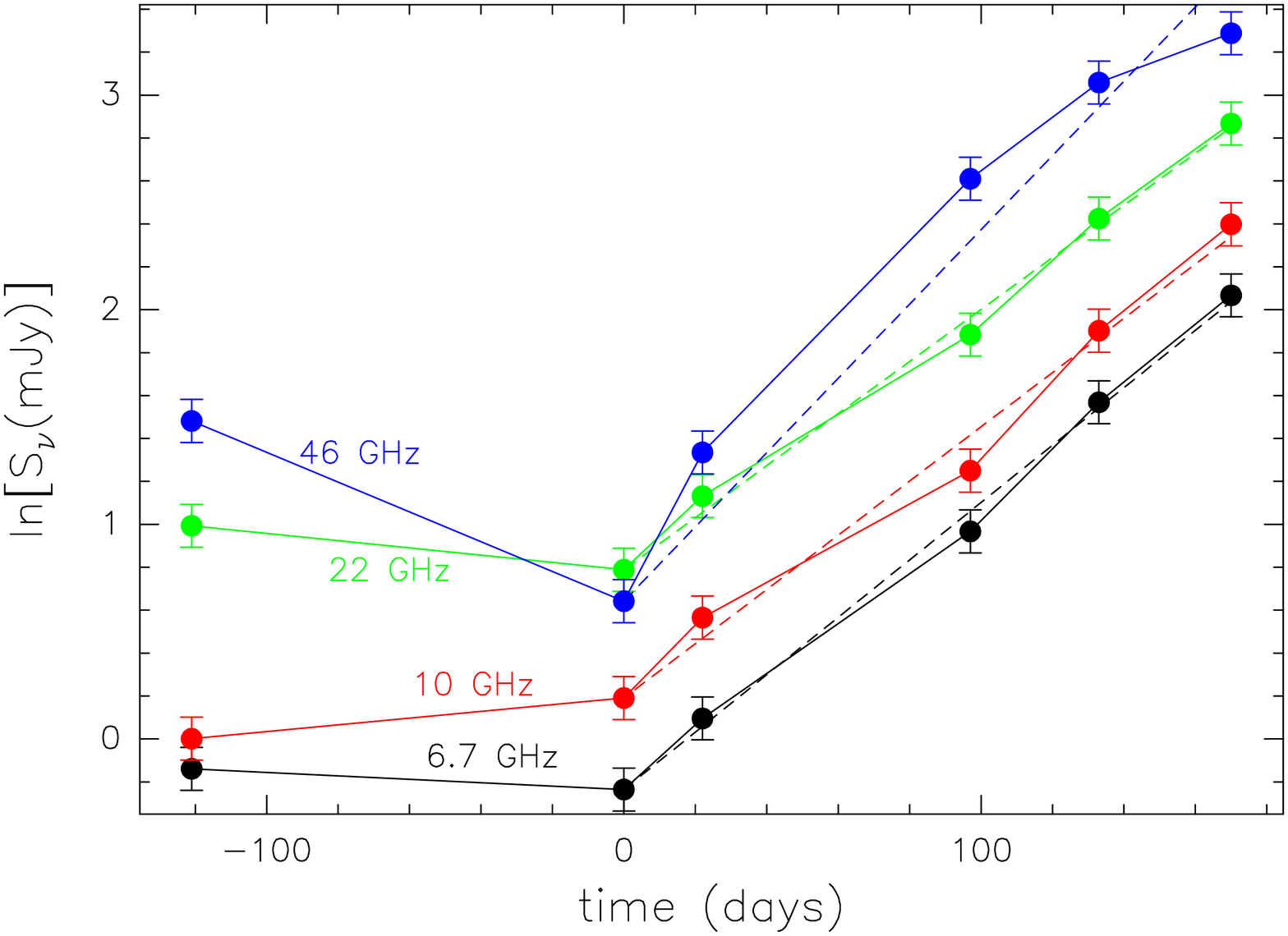}}
\caption{
Observed radio flux densities of NIRS\,3 as a function of time, with $t$=0
corresponding to the second epoch of our monitoring (July 10, 2016). The
solid lines connect fluxes observed at the same frequency, which is
given beside each curve.  The dashed lines are the fits assuming
exponential dependence on $t$ according to Eq.~(\ref{eft}).
}
\label{fft}
\end{figure}

Figure~\ref{fft} illustrates the evolution of the flux density of NIRS\,3
from March to December 2016. As already noted, after an initial
quiescent phase, the flux rose between July and August, undergoing an
exponential increase. The dashed lines in the figure are fits to the
expressions of the flux density, $S_\nu$, as a function of time, $t$,
\begin{equation}
 \frac{S_\nu}{S_\nu(t_0)} = \exp\left(\frac{t-t_0}{T}\right)   \label{eft}
,\end{equation}
where we assume $t_0=0$ on July~10 and the timescale $T$ is the
free parameter of the fit. The value of $T$ ranges from 58~days (for the
45~GHz data) to 82~days (for the 22~GHz data), with a mean value of 77~days.

Clearly, the fits in Fig.~\ref{fft} indicate that the radio burst has
indeed started very close to t=0, i.e. July 10, 2016. Therefore, we assume
that the increase in the radio flux  became detectable right after this date.

Assuming that the radio burst is due to expansion of the jet on a timescale equal to $T$, with a typical expansion speed of 660~\kms\ (see above), we find a
corresponding spatial scale of $\sim$30~au. While this value has no precise
correspondence with a specific geometrical feature of the jet,
nonetheless it indicates that the jet is very compact and arises close to
the circumstellar disk.

We can also use the expansion speed to derive a rough estimate of the
maximum jet size, at the most recent observing epoch (December 27, 2016).
With a time lag of $\sim$560~days from the estimated beginning of the
outburst (mid-June 2015), the jet should have expanded up to $\sim$210~au or
0\farcs12 (assuming that accretion and ejection occur at the same time).

\begin{figure}
\centering
\resizebox{8.5cm}{!}{\includegraphics[angle=0]{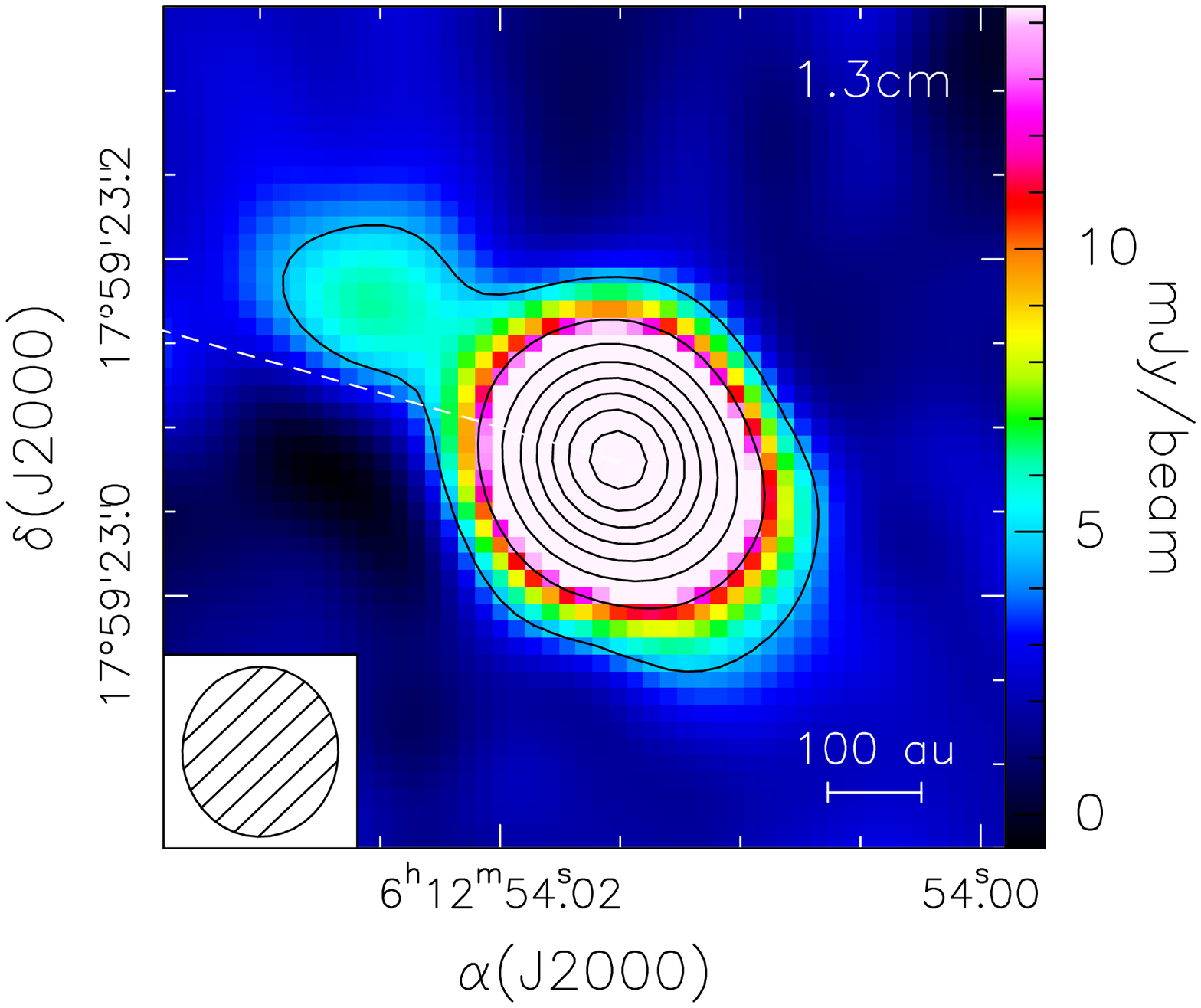}}
\caption{
Continuum map at 22.2~GHz of the radio source NIRS\,3 obtained at the fifth epoch of our
monitoring (December~27, 2016). The image has been saturated to emphasize
the elongated structure. The dashed line denotes the direction connecting
NIRS\,3 to the NE knot of the large-scale radio jet. 
Contour levels range from 0.6 (7$\sigma$) to 13.2 in steps of 1.8~mJy/beam.
The ellipse
in the bottom left denotes the synthesized beam.
}
\label{felo}
\end{figure}

A jet of this  size should be barely resolved on our maps at the highest
frequencies. Indeed, at 22.2~GHz we do reveal an elongated structure just in
the direction of the NE knot of the larger scale radio jet. Figure~\ref{felo}
shows the corresponding map, where the low-level emission has been suitably
enhanced. The approximate size on the plane of the sky is $\sim$300~au.  We note that
the same structure is not detected at 7~mm
because at this band the uv coverage and hence the corresponding image
quality has been significantly degraded due to heavy flagging of the data.
This finding provides us with direct evidence of the existence of
an expanding jet on scales of a few 100~au.

\subsection{Millimetre emission}

Our monitoring with the IRAM/NOEMA and ALMA interferometers provides
flux density measurements starting from December 16, 2016. In addition to this
observation, we  report on two more epochs, namely January 23 and
February 15, 2017. These data allow us to establish that the millimetre
emission from source NIRS\,3 is basically constant in time.  Our angular
resolution is sufficient to resolve the hot molecular core associated with
NIRS\,3, named SMA1 by Zinchenko et al.~(\cite{zin12}), from the nearby
fainter core, named SMA2 by the same authors. In Table~\ref{tmon} we give
the mm flux densities of NIRS\,3 at the different epochs. Clearly, no
change is seen within a calibration error of $\sim$20\%.

\begin{table}
\centering
\caption[]{
Flux densities measured at mm wavelengths towards \S\ with the SMA by various
authors.
}
\label{tfmm}
\begin{tabular}{cccc}
\hline
\hline
$\nu$ & HPBW & $S_\nu$ & reference \\
(GHz) & (arcsec) & (mJy) & \\
\hline
225   & 0.45 &  58 & Zinchenko et al.~(\cite{zin15}) \\
225   & 3.37 & 290 & Zinchenko et al.~(\cite{zin12}) \\
230.5 & 1.35 & 171 & Wang et al.~(\cite{wang}) \\
284   & 2.75 & 450 & Zinchenko et al.~(\cite{zin12}) \\
350   & 2.04 & 500 & Zinchenko et al.~(\cite{zin15}) \\
\hline
\hline
\end{tabular}
\end{table}

As already done for the VLA data, it is worth  comparing our
flux density measurements with those from the literature. However,
unlike the case of the centimetre emission, direct comparison with the
literature values is impossible because they were  taken at
shorter wavelengths. Moreover, a variety of array configurations were
used, which further complicates the comparison of literature fluxes from
different authors.

In an attempt to overcome this problem, we have collected all the
results from observations of \S\ performed with the same instrument, the
Submillimeter Array (SMA). These are given in Table~\ref{tfmm}, with
the observing frequency and angular resolution. The large spread of
the flux densities at the same frequency is due to the  more
extended array configurations which provide better angular resolutions, but also
resolve out part of the emission. To correct for this effect, we assume
that the largest angular scale, $\Theta_{\rm L}$, that can be imaged by
the interferometer is proportional to the synthesized beam, $\Theta_{\rm B}$.
This  assumption is reasonable for the given interferometer. We also
assume that the flux density measured by the interferometer at a given
frequency towards a given source scales as
\begin{equation}
 S_\nu \propto \Theta_{\rm L}^\alpha \propto \Theta_{\rm B}^\alpha   \label{eth}
\end{equation}
with $\alpha>0$. This empirical relation expresses the concept that
the lower  the resolution, the smaller the amount of flux resolved out by the
interferometer.

\begin{figure}
\centering
\resizebox{8.5cm}{!}{\includegraphics[angle=0]{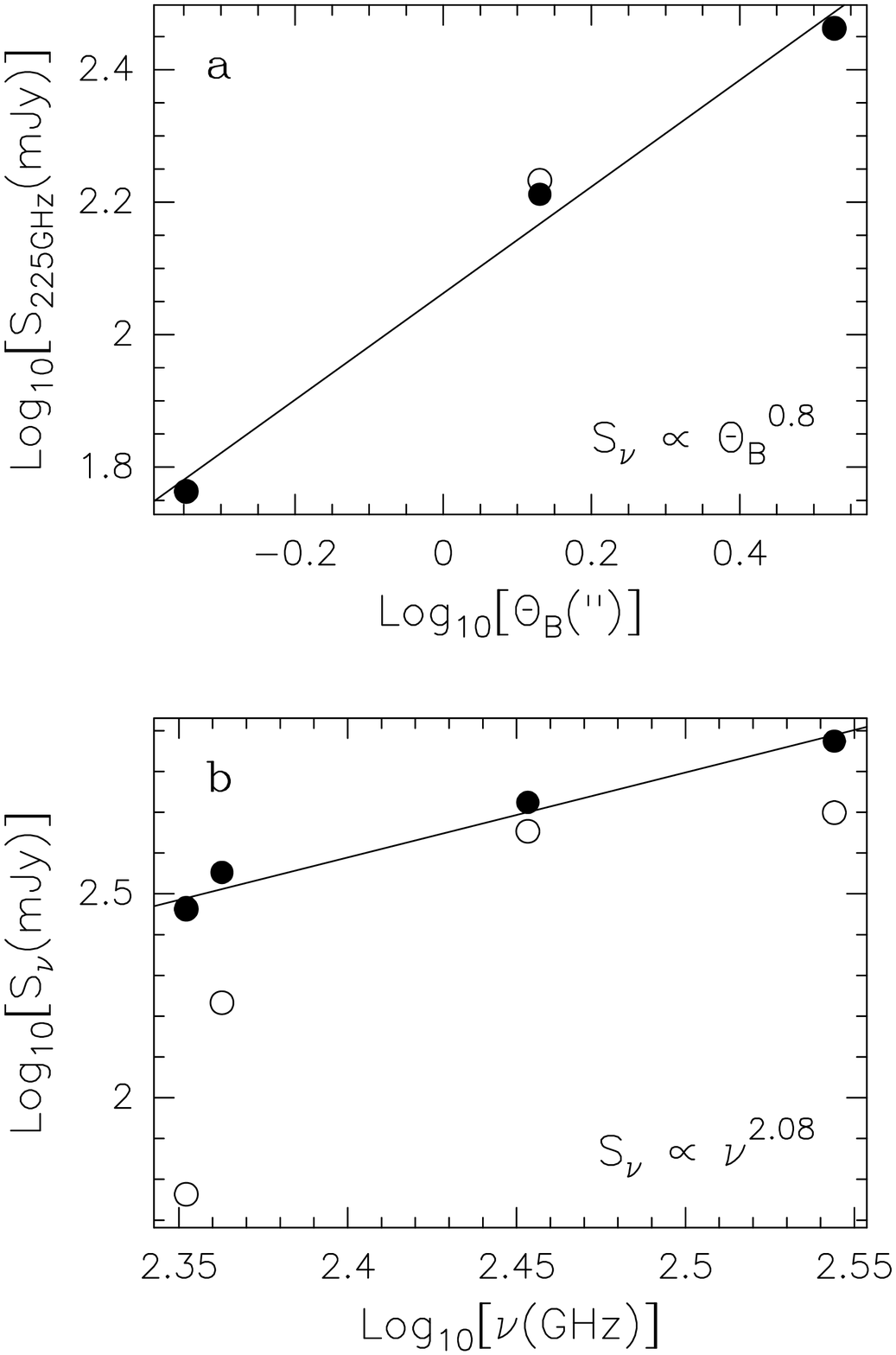}}
\caption{
{\bf a.} Flux densities at 225~GHz and 230~GHz from Table~\ref{tfmm}
versus the full width at half power of the corresponding synthesized
beam. The 230~GHz flux (empty point) has been scaled to 225~GHz (solid
point) assuming $S_\nu\propto\nu^2$. The straight line is the linear fit
(in the logarithmic plot) to the solid points, based on Eq.~(\ref{efit}).
{\bf b.} Flux densities from Table~\ref{tfmm} as a function of
frequency before (empty points) and after (solid points) scaling the fluxes
to an angular resolution of 3\farcs37 (see text) with Eq.~(\ref{efit}).
The straight line is the linear fit (in the logarithmic plot) to the solid
points, based on Eq.~(\ref{etot}).
}
\label{fmmcorr}
\end{figure}

To determine $\alpha$ we have used the fluxes from the literature measured
at the same frequency with different resolutions. In Fig.~\ref{fmmcorr}a, we
plot the flux density versus the half-power width of the synthesized beam for
all measurements from Table~\ref{tfmm} at 225~GHz. We have also included the
230~GHz flux density from Wang et al.~(\cite{wang}) after scaling it to
225~GHz, assuming a spectral index\footnote{A different spectral index, e.g. 4, does not
change the estimate of $\alpha$ significantly.} of 2. Fitting Eq.~(\ref{eth}) to this
plot, we obtain the expression
\begin{equation}
 S_{\rm 225\,GHz}({\rm mJy}) = 115.5 \, \left[\Theta_{\rm B}({\rm arcsec})\right]^{0.8}.
 \label{efit}
\end{equation}
This expression applies only to the range of fluxes and $\Theta_{\rm B}$
in Fig.~\ref{fmmcorr}a. Any extrapolation beyond these limits or application
of Eq.~(\ref{efit}) to interferometers other than the SMA is unreliable.

Using this equation,
we can  scale all fluxes in the table to the same
angular resolution. We adopt the lowest resolution in Table~\ref{tfmm},
$\Theta_{\rm B}=3\farcs37$ because this is the closest to the
resolution of our ALMA and IRAM/NOEMA images and  likely recovers
most of the flux from the NIRS\,3 hot molecular core.

From the corrected fluxes we can
then derive the spectral slope $S_\nu\propto\nu^\beta$. This is done
in Fig.~\ref{fmmcorr}b, where the empty points indicate the original
fluxes measured with the SMA and the solid points those after scaling for
$S_\nu\propto\Theta_{\rm B}^{0.8}$. The best fit to the corrected fluxes is
\begin{equation}
 S_\nu({\rm mJy}) = 3.89\times10^{-3} \, \left[\nu({\rm GHz})\right]^{2.08}.
 \label{etot}
\end{equation}

This expression allows extrapolation to the bands covered by our observations
and comparison to the corresponding flux densities. The extrapolated fluxes
at 2~mm and 3~mm are 20--40\% less than those measured by us. Even allowing
for a 20\% calibration error, this  difference suggests that the millimetre
source could be now slightly brighter than in the pre-burst phase. We 
discuss this possibility in Sect.~\ref{smm}.

\section{Interpretation: An expanding jet model}
\label{sdis}

We wish to demonstrate that an expanding thermal jet boosted by the accretion
outburst is a viable explanation for the observed exponential increase in
the radio emission from \S.

From the archival data we know that the radio jet was already present
at the moment of the radio burst, which means that any realistic jet model should consider
a scenario where a pre-existent thermal jet is suddenly perturbed by a
massive ejection event. However, modelling the propagation of a similar
perturbation across the jet is too complicated for our purposes. We thus
prefer to simplify the problem by assuming that the expanding jet due to
the outburst becomes detectable only when its radio flux overcomes that of the
pre-existent jet. While this is not strictly correct, it should nevertheless
provide us with an acceptable description of the outbursting component of
the jet, which is what we actually measure with our VLA monitoring.

In the following, we adopt the jet model by Reynolds~(\cite{reyn}). While
this model considers a number of possible jet shapes and distributions
of the relevant physical parameters (ionization fraction, density,
expansion velocity, etc.), in order to simplify the problem we  adopt
the simplest representation, corresponding to the ``standard spherical''
case in Reynolds' Table~1.  This consists of a conical jet where the opening
angle (\tho), electron temperature ($T_0$), expansion speed ($\varv_0$),
and ionization fraction ($x_0$) do not depend on the distance from the star,
$r$, while the ionized gas density is $\propto r^{-2}$.

Following Reynolds' notation, and in particular his Eqs.~(9) and~(10), the
total flux density at frequency $\nu$ from both jet lobes\footnote{ Reynolds' Eq.~(8) takes into account emission from only one lobe of
the jet, whereas we assume the jet to be bipolar and multiply the flux by
a factor of 2 to take both lobes into account.} is given by the following expression:
\begin{eqnarray}
S_\nu & = & \frac{4\,a_{\rm j}}{d^2\,a_\kappa} \theta_0 r_0 T_0 \nu^2 r_0 \sin\,i \nonumber \\
& & \times \left\{ 
\begin{array}{lcl}
\tau_0 \left(1-\frac{y_0}{y_{\rm max}}\right) & \Leftrightarrow & y_1<y_0 \\
\frac{1}{2} \left[\left(\frac{y_1}{y_0}\right)^2-1\right] +
            \tau_0 \left(\frac{y_0}{y_1}-\frac{y_0}{y_{\rm max}}\right) & \Leftrightarrow & y_0<y_1<y_{\rm max} \\ 
\frac{1}{2} \left[\left(\frac{y_{\rm max}}{y_0}\right)^2-1\right] & \Leftrightarrow & y_1>y_{\rm max}
\label{eflux}
\end{array}
\right. 
\end{eqnarray}
If all quantities are expressed in CGS units,
$a_{\rm j}$=$6.50\times10^{-38}$ and $a_\kappa$=0.212 (see Reynolds'
Eqs.~(2) and~(3)).  Here, $y$ is the impact parameter of a generic line of
sight, i.e. $y=r\sin\,i$, with $i$ the angle between the jet axis and the line
of sight; $r_0$ is the minimum distance of the jet from the star (namely
the radius at which the ionized material is injected); $r_{\rm max}$
the maximum extension of the jet; $y_1$ corresponds to opacity $\tau=1$;
and $\tau_0$ is the opacity along the line of sight with impact parameter
$y_0=r_0\,\sin\,i$. From Reynolds' Eqs.~(4) and~(12) and his expression of
the total (ionized plus neutral) mass loss rate, $\dot{M}$, we can obtain
the following expressions for $\tau_0$ and $y_1$:
\begin{eqnarray}
\tau_0 & = & \frac{2\,a_\kappa}{(\pi\mu)^2}\,(\theta_0 r_0)^{-3} \left(\frac{x_0\,\dot{M}}{\varv_0}\right)^2 T_0^{-1.35} \nu^{-2.1} (\sin\,i)^{-1} ,\\
y_1 & = & \left[\frac{2\,a_\kappa}{(\pi\mu)^2}\,\theta_0^{-3} \left(\frac{x_0\,\dot{M}}{\varv_0}\right)^2 T_0^{-1.35} \nu^{-2.1} (\sin\,i)^2\right]^\frac{1}{3}
.\end{eqnarray}
Here, $x_0$ is the fraction of ionized gas and $\mu$ the mean particle
mass per hydrogen atom.

As illustrated in Appendix~\ref{sapp},
the typical spectrum obtained from Eq.~(\ref{eflux}) is characterized
by optically thick emission at low frequencies, growing as $\nu^2$,
and optically thin emission at high frequencies, slowly decreasing
as $\nu^{-0.1}$. Between these two regimes, the combination of thick
and thin emission results in a spectral index of $\sim$0.6 (see Reynolds' Table~1).

\subsection{Determining the best-fit model}
\label{sbf}

Our purpose is to show that the observed flux variations can be fitted with
this model and thus derive a number of important physical parameters of the
jet. To reduce the number of free parameters, we fix the electron temperature
to $T_0=10^4$~K, a typical value for ionized gas associated with YSOs.

We adopt a simplified scenario, where the gas becomes ionized just when it
reaches the terminal velocity, $\varv_0$, at $r=r_0$. We also assume that
the ejected material undergoes constant acceleration starting from $r=0$
up to $r=r_0$, and beyond this radius expands at constant speed. The time
$t_0$ that is needed to reach $r_0$ with constant acceleration from zero
velocity is $t_0=2\,r_0/\varv_0$. The maximum radius of the jet at time
$t\ge t_0$ can be written as
\begin{equation}
 r_{\rm max} = \varv_0\,(t-t_0)+r_0 = \varv_0\,t-r_0
\end{equation}
After multiplying both members by $\sin\,i$, we obtain the expression
for the jet expansion in the plane of the sky,
\begin{equation}
 y_{\rm max} = \varv_0'\,t-y_0   \label{eexp}
,\end{equation}
where we have defined ~$\varv_0'=\varv_0\,\sin\,i$. We note that the time
$t$ is measured from the beginning of the outburst (mid-June 2015), under the
assumption that the accretion and ejection bursts occur at the same time.

While angles between 80\degr\ (Boley et al.~\cite{bole}) and 20\degr\
(Zinchenko et al.~\cite{zin15}) have been proposed for the inclination
of the jet, we prefer to adopt $i=80\degr$ because a jet directed close
to the line of sight appears inconsistent with Fig.~1 of Caratti o Garatti et
al.~(\cite{cagana}). This figure shows that scattered IR emission is clearly
detected from both lobes, whereas for $i$ close to 0 only one lobe should
be seen.

The expansion speed of the jet can be estimated from a spectrum of
the Paschen $\beta$ recombination line obtained by Caratti o Garatti
(priv. comm.), which presents a P Cygni profile with a broad absorption
feature extending up to $\sim$900~\kms, as well as a wide red-shifted wing with similar velocity. Since this is probably
seen through scattered light along the outflow lobes, this velocity should
be very close to the true expansion speed of the ionized jet.

We can constrain the value of $r_0$ using the observed shapes of
the continuum spectra (see Fig.~\ref{fbest}).  In fact, we can demonstrate
(see Appendix~\ref{sapp}) that the ratio $r_0/r_{\rm max}$ depends only
on the ratio between the two turn-over frequencies characterizing the
typical spectrum of a radio jet. From Fig.~\ref{felo} we find that
in December~2016 the maximum jet extension was $y_{\rm max}\la300$~au,
while from Fig.~\ref{fbest} we see that the two turn-over frequencies are
$\nu_{\rm m}>45.5$~GHz and $\nu_{\rm t}<6$~GHz, hence from Eq.~(\ref{eqap})
we obtain $r_0<70$~au.
Furthermore, we know that the radio burst occurred
$t_{\rm lag}\simeq13$~months after the IR outburst. This time lag should
correspond to the time interval needed for the jet to be ionized and
expand until its flux overcomes the pre-existent radio flux. This implies
that $t_{\rm lag} \ga t_0=2\,r_0/\varv_0$, namely $r_0\la100$~au.

In conclusion, $r_0$ must be less than $\sim$70~au, but not much less,
otherwise the value of $t_0$ would be too small with respect to the observed
time lag between the IR and radio bursts (see also Sect.~\ref{sro}).
We believe that $r_0=50$~au is a reasonable compromise because theoretical
jet models indicate that the terminal speed is attained at a few au from
stars of a few $M_\odot$ (see Pudritz et al.~\cite{pudppv} and references
therein). It is thus plausible that such a distance may scale up by an
order of magnitude for stars 10 times more massive.

From Eq.~(\ref{eexp}) we can now estimate the values of $y_{\rm max}$ at the
five epochs of the radio burst. Starting from July 10, 2016, these are 150,
161, 199, 218, and 237~au.

Having fixed $r_0$, $i$, $T_0$, and \ymax\ at each epoch, we are left
with only two free parameters: \tho, and $x_0\,\dot{M}/\varv_0$.  To obtain
the best fit to the observed fluxes, we have varied these two parameters
over a suitable grid and identified the pair of values which minimizes
the expression
\begin{equation}
\chi^2 = \sum\limits_{i=1}^4 \left[\frac{\Log(S_{\nu_i}^{\rm obs})-\Log(S_{\nu_i}^{\rm mod})}{\sigma}\right]^2
\label{echi}
,\end{equation}
where $S_{\nu_i}^{\rm obs}$ is the flux density measured at each observing
frequency $\nu_i$, $S_{\nu_i}^{\rm mod}$ the model flux density obtained
from Eq.~(\ref{eflux}), and ~$\sigma=0.1~\Log\,\e$~ is the error on
$\Log(S_{\nu_i}^{\rm obs})$, corresponding to 10\% calibration uncertainty
at all VLA bands.

\begin{figure*}
\centering
\resizebox{17.5cm}{!}{\includegraphics[angle=0]{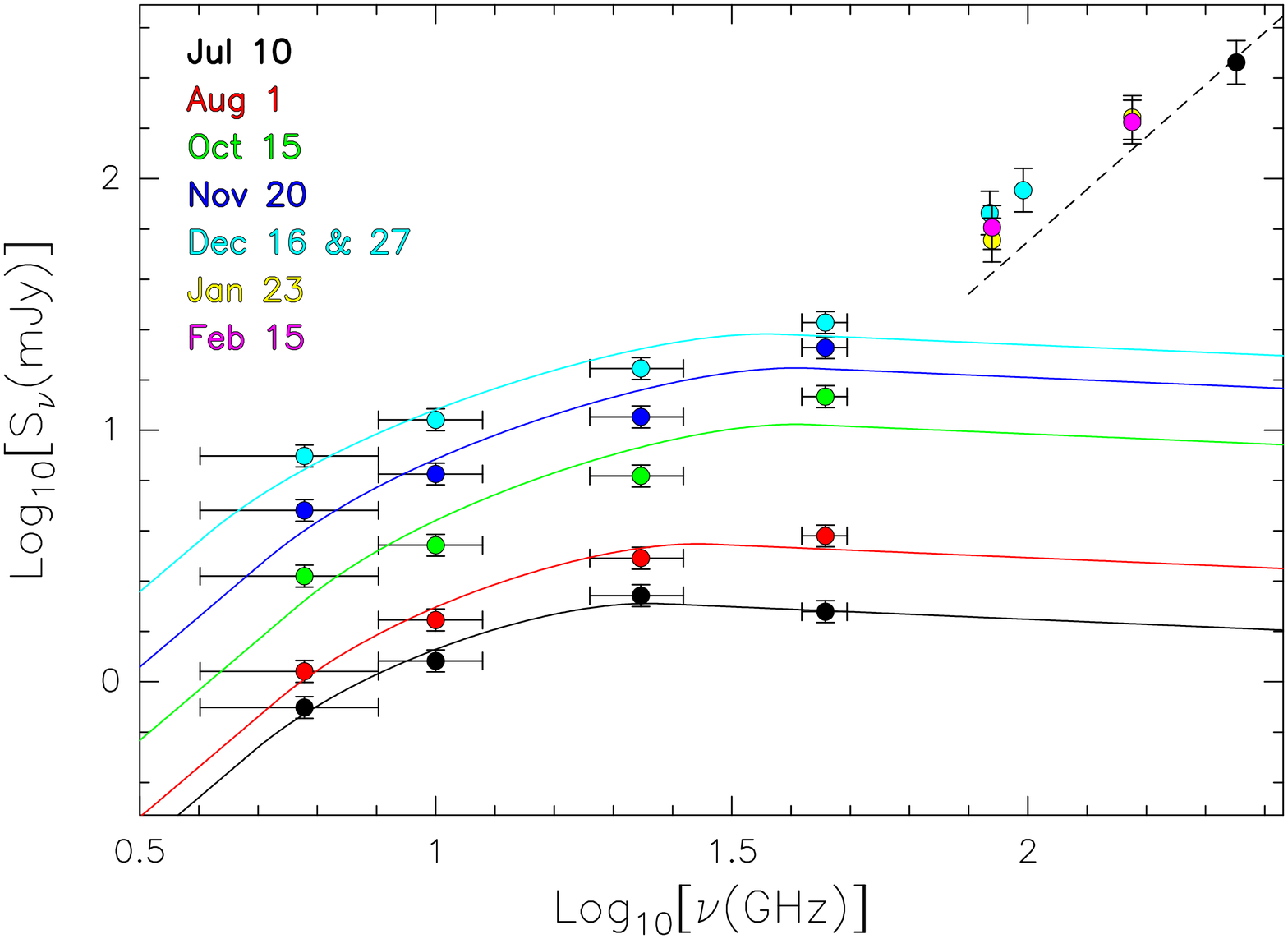}}
\caption{
Flux density measured from \S\ as a function of frequency (points). The
curves represent the best fits to the different epochs obtained with the
model by Reynolds~(\cite{reyn}).
Symbols with the same colour correspond to the
same epoch, as explained in the figure.  The black point at 225~GHz is the
measurement by Zinchenko et al.~(\cite{zin12}).  The dashed line denotes the
expected mm emission according to Eq.~(\ref{etot}).  The vertical error bars
correspond to the calibration uncertainty, while the horizontal bars indicate
the frequency ranges covered by the VLA correlator at the different bands.
}
\label{fbest}
\end{figure*}

The best-fit is represented by the curves in Fig.~\ref{fbest}, which satisfactorily
fit the data points (solid circles with the same colour as the curve). In Table~\ref{tbf} we give the parameters for which the
best fit has been obtained at each epoch. We stress that changing our
assumption of $r_0=50~$au by about $\pm$10~au would not affect the quality
of the fits significantly.

\begin{table}
\centering
\caption[]{
Best-fit parameters at each observing epoch.
}
\label{tbf}
\begin{tabular}{cccc}
\hline
\hline
date & $\theta_0$ & $x_0\,\dot{M}/\varv_0$ \\
     & (deg) & ($10^{-9}~M_\odot$~yr$^{-1}$/(\kms)) \\
\hline
July 10     & 14$\pm$2.4 & 1.2$\pm$0.21 \\
August 1    & 16$\pm$2.4 & 1.8$\pm$0.30 \\
October 15  & 20$\pm$2.7 & 3.8$\pm$0.56 \\
November 20 & 32$\pm$5.0 & 7.9$\pm$1.3 \\
December 27 & 54$\pm$11 & 15$\pm$2.9 \\
\hline
\hline
\end{tabular}
\end{table}

\subsection{Origin of the millimetre emission}
\label{smm}

Thanks to our model fit, we can also analyse the nature of the millimetre
emission observed with NOEMA and ALMA.  It is interesting to note (see
Fig.~\ref{fbest}) that the flux densities measured with ALMA at 86.2 and
98.2~GHz (cyan points; see Table~\ref{tmon}) are comparable, within the
errors, to the sum of the free-free emission from the jet measured at the
same epoch (cyan curve), i.e. $\sim$22~mJy, plus the dust contribution of
41~mJy at 86.2~GHz and 54~mJy at 98.2~GHz expected at these frequencies on
the basis of Eq.~(\ref{etot}) (dashed line). We conclude that the small
flux increase observed at millimetre wavelengths with respect to the
(extrapolated) pre-burst literature flux can be entirely due to the free-free
emission from the radio jet and not to dust heating by the outburst.

This result differs from the (sub)mm brightening detected by Hunter
et al.~(\cite{hunt}) in a similar high-mass YSO (NGC6334I-MM1), also
undergoing a maser flare and, possibly, an accretion outburst.  Such a
lack of brightening in the (sub)mm regime is puzzling given the
similarities between the two outbursts. However, in our case the increase in
the bolometric luminosity was $\sim$13 times less than for NGC6334I-MM1,
and \S\ appears to be less deeply embedded. These differences could
justify a shorter timescale for the decay of the dust temperature in our
object. Na\"ively, we might expect that a burst of energy takes longer to be
dissipated in a thick envelope than in a disk for one simple geometrical
reason: the surface-to-volume ratio of a thin cylinder is much greater
than that of a sphere with the same radius, thus making irradiation more
effective in cooling the gas.

\subsection{Time lag between the IR and radio bursts}
\label{sro}

As seen in Sect.~\ref{sbf}, the distance from the star at which the radio
jet becomes ionized must be $r_0<70$~au.  Such a small value of $r_0$ implies a
timescale $t_0=2\,r_0/\varv_0<9$~months for the ejected material travelling
at $\varv_0=900$~\kms\ to reach $r_0$ and trigger the jet ionization. If
the onset of the ejection episode coincides with the accretion outburst
detected in the IR, $t_0$ appears inconsistent with the observed time
lag between the IR and the radio bursts, which is significantly longer
($\sim$13~months). However, it should be kept in mind that the brightening
of the radio emission can be detected only when the intensity of the radio
burst overcomes the pre-existent flux density. For this to happen, it is
necessary for the jet to expand well beyond $r_0$ and it is hence plausible
that a few more months are needed,  corresponding to a jet expansion  of
the order of a few 10~au.  In conclusion, we believe that the 13-month
lag that we observed  is compatible with the value of $r_0=50$~au adopted in
our model.

A consequence of the observed time lag is that the radio burst 
is unlikely to
be
due to photoionization because in this case the burst should have been
detected at the same time at both IR and radio wavelengths.
In fact, the timescale to photoionize the gas can be estimated using
Eqs.~(5.6) and (7.7) in Dyson \& Williams~(\cite{dywi}). Assuming a
density of $\sim$$6\times10^8$~\cmc\ for the inner regions of the
disk (see Zinchenko et al.~(\cite{zin15})), we find a timescale of only
$\sim$2.3~hours, much less than the 13-month time lag between the IR
burst and the radio burst. While it is possible that the density is lower
in the direction of the disk axis, it seems more likely that the radio
burst is triggered mechanically rather than by a ``flash'' of light.
This scenario is consistent with a low ionization degree of the ejected gas,
as discussed later in Sect.~\ref{sxo}.

\subsection{Evolution of jet parameters}

\begin{figure}
\centering
\resizebox{8.5cm}{!}{\includegraphics[angle=0]{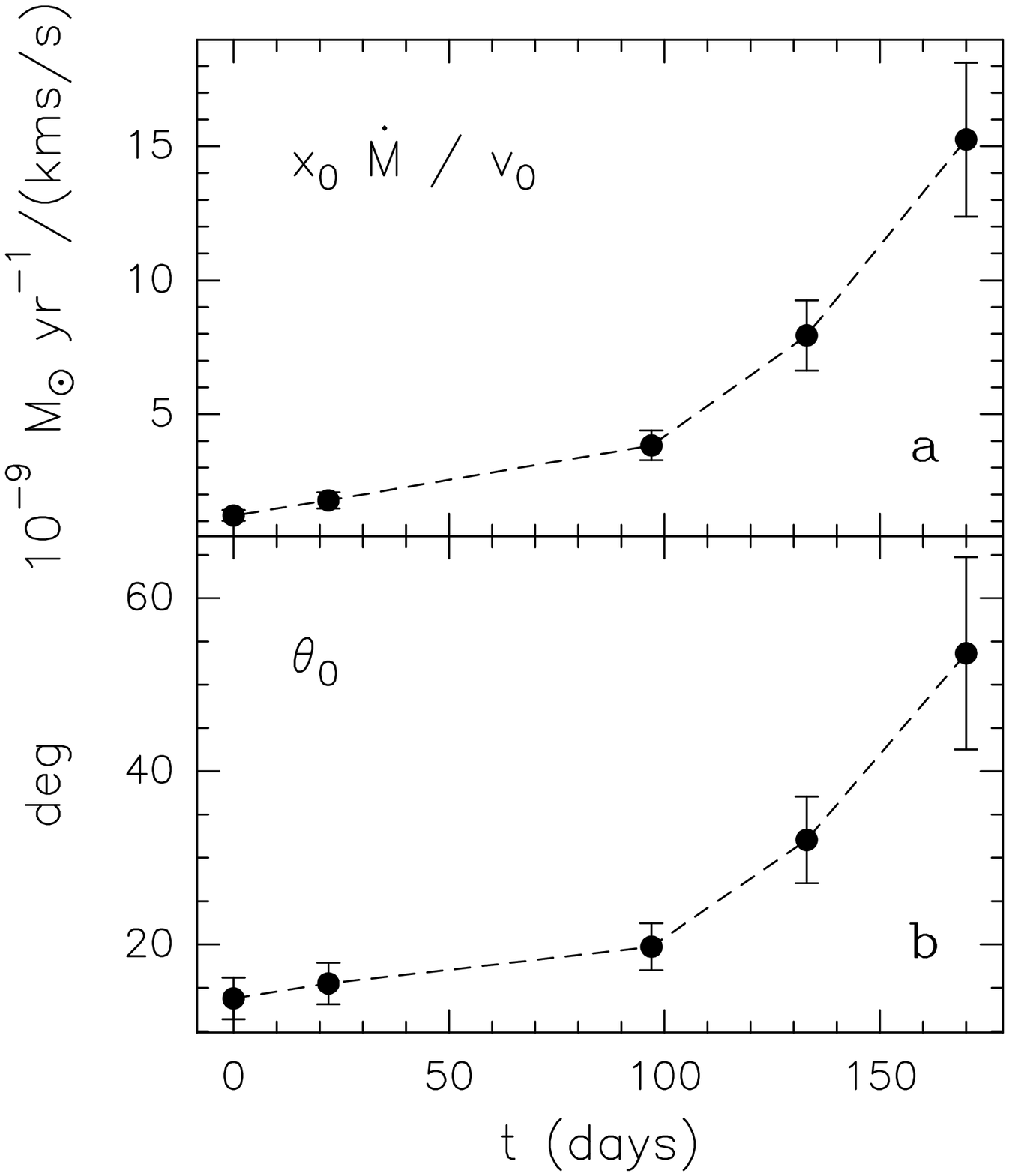}}
\caption{
{\bf a.} Plot of the best-fit parameter $x_0\,\dot{M}/\varv_0$ at the five
observing epochs. The initial time, $t$=0, corresponds to July~10, 2016.
{\bf b.} Same as the previous panel, but for the other best-fit parameter $\theta_0$.
}
\label{fpars}
\end{figure}

The fits to the observed spectra allow us to investigate the variation
of the jet physical parameters with time. Figures~\ref{fpars}a and~\ref{fpars}b
show the best-fit values of $x_0\,\dot{M}/\varv_0$ and \tho\ as a function
of time from the estimated beginning of the radio burst. The errors were computed
by applying the method used by Lampton et al.~(\cite{lamp}) to the expression of
$\chi^2$ in Eq.~(\ref{echi}), for two free parameters.

From Table~\ref{tbf} we see that at
the beginning of the radio burst the jet is quite collimated
and the ionized mass loss rate is lower,
of the order of $10^{-6}~M_\odot$~yr$^{-1}$. Subsequently, we see a
steady increase in \tho,\ which indicates that the jet is progressively
widening, while the ejection rate of the ionized gas $x_0\,\dot{M}$ is
increasing. This result might look surprising because, according to
our ongoing monitoring at IR wavelengths, the peak of the IR light-curve
lies between November and December~2015; therefore, the outburst in December
2016 should  already be fading. However, a few considerations are in order.

\begin{figure}
\centering
\resizebox{8.5cm}{!}{\includegraphics[angle=0]{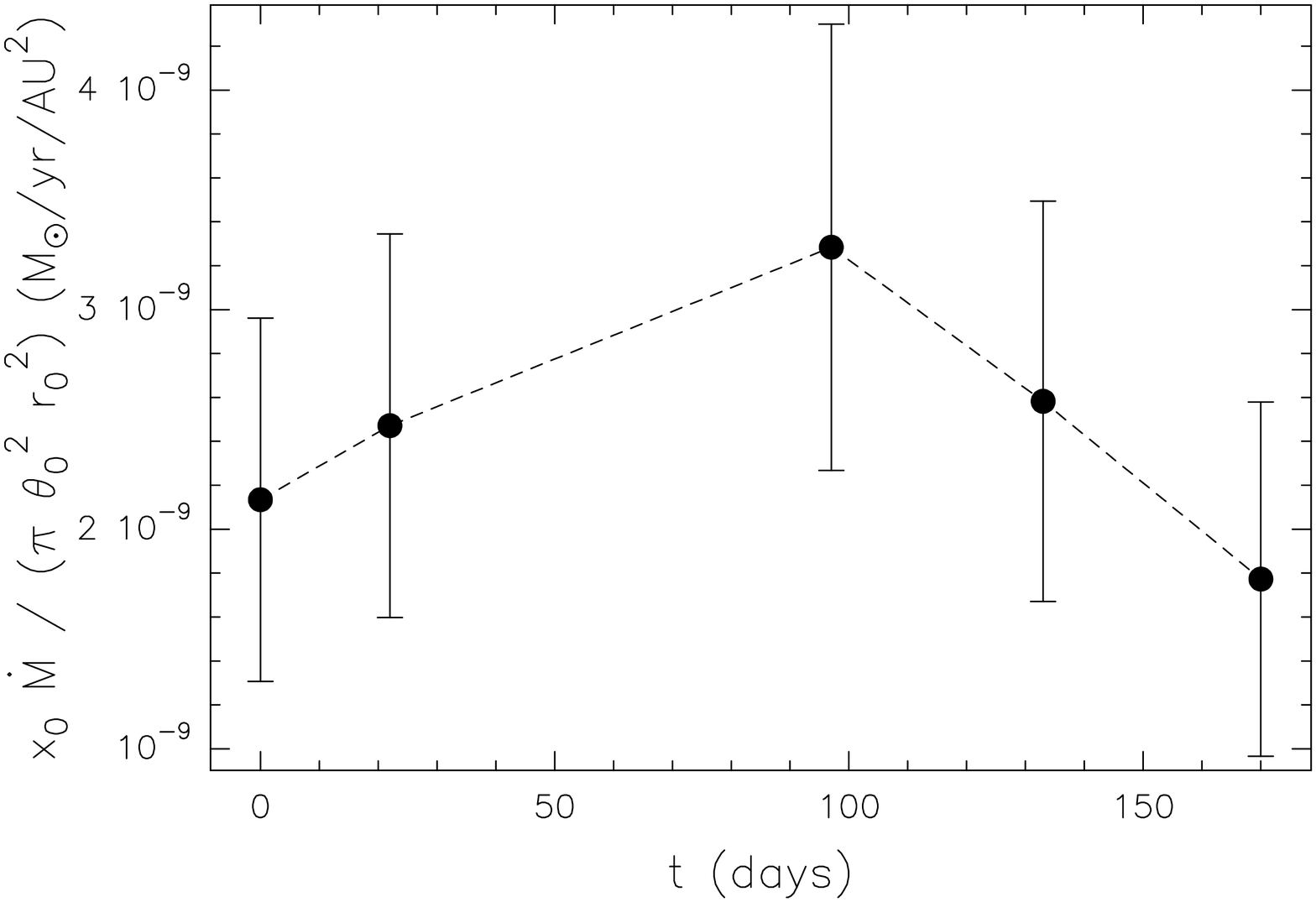}}
\caption{
Ionized mass flux through the radio jet at the five observing epochs.
No significant change is observed within the errors.
}
\label{fno}
\end{figure}

First of all, we do not have a precise estimate of the ionized fraction
$x_0$ and can only set an upper limit (see Sect.~\ref{sxo}). Therefore, the
increase in $x_0\,\dot{M}$ does not necessarily correspond to an increase
in the mass loss rate, $\dot{M}$.
Furthermore, a rise in $x_0\,\dot{M}$ can be explained by
the increase in \tho\ because the ionized mass flux at $r_0$,
$x_0\,\dot{M}/(\pi\theta_0^2 r_0^2)$ is constant within the errors, as
shown in Fig.~\ref{fno}. Therefore, the augmented ejection rate of the
ionized gas could be due to a widening of the ejection angle.

Finally,  we actually do know that the radio burst occurred with
a delay of $\sim$13~months after the onset of the accretion burst and it is
hence reasonable that  the fading of the emission at the two bands
also presents such a delay. This would imply that the radio intensity should begin
to decrease at the beginning of 2017, $\sim$13~months after the peak of the
IR burst, which falls between December~2015 and January~2016 (Caratti
o Garatti priv. comm.). It is thus not surprising that the radio flux
measured in December~2016 is still rising if the light curve of the radio
burst mimicks the evolution of the accretion outburst with a delay of 13~months.

\subsection{Weakly ionized jet}
\label{sxo}

For the first time our study makes it possible to compute the
ionization degree of a thermal jet, thanks to our knowledge of both
the mass accretion rate and the ionized mass loss rate. Caratti o
Garatti et al.~(\cite{cagana}) have estimated an accretion rate of
at least $\sim5\times10^{-3}~M_\odot~$yr$^{-1}$ during the outburst. It is reasonable
that at least 10\% of the infalling material will be redirected into
the outflow, so we can assume $\dot{M}>10^{-4}$~\Msun~yr$^{-1}$. From
$x_0\,\dot{M}/\varv_0<1.5\times10^{-8}~M_\odot$~yr$^{-1}/$(\kms)
(see Table~\ref{tbf}) and $\varv_0\simeq900$~\kms, we obtain
$x_0\,\dot{M}<1.4\times10^{-5}$~\Msun~yr$^{-1}$ and $x_0<0.14$.
We note that this is a conservative upper limit. This result
demonstrates that the ionized component is only a small fraction of the
whole jet, which is mostly neutral.

Indeed, comparison with other studies strongly suggests that a low degree of
ionization could be a common feature of all radio jets. For example,
Guzm\'an et al.~(\cite{guz10}), Purser et al.~(\cite{purs}), and
Sanna et al.~(\cite{sanna16}) find ionized mass loss rates ranging from
$2\times10^{-7}$ to $8\times10^{-6}$~\Msun~yr$^{-1}$, values even lower
than that estimated by us for \S.
In conclusion, our results demonstrate that the ionized component of the
jet is dynamically negligible.

Finally, we note that a jet with a low ionization fraction is
clearly different from the fully ionized bubbles predicted by Tanaka et
al.~(\cite{tana}). Since in their case the gas is photoionized, our finding
further supports the idea that in \S\ we are observing a thermal jet.
In turn, the lack of an \HII\ region around a (proto)star as massive as \S\
($\sim$20~\Msun; Zinchenko et al.~\cite{zin15}) suggests that such
a region could be quenched, consistent with the large accretion rate
estimated in our case (see Yorke~\cite{yorke86}, Walmsley~\cite{walms}). In
addition, we could be dealing with a bloated, colder protostar, whose
Lyman continuum emission is much less than that expected for a zero-age
main-sequence star with the same mass (see Hosokawa et al.~\cite{hoso10}).

\section{Summary and conclusions}
\label{scon}

We have reported on the first detection of a radio burst in a massive YSO
(\S) that had previously undergone an accretion outburst revealed in the
methanol maser and IR emission. Along with methanol maser flares, IR and
mm flux variability, the radio variability from thermal jets can thus be 
considered  an additional signpost of accretion bursts in high-mass YSOs,
especially for deeply embedded objects where even the near-IR emission can be
heavily extincted.

The structure of the radio source seen in archival VLA maps prior to the
burst is strongly suggestive of a bipolar thermal jet, with two lobes
extending NE and SW. The burst is observed only towards the central,
unresolved radio source presumably coinciding with the YSO.

The analysis of the first six epochs of our on-going monitoring in the range
6--46~GHz indicates that the radio burst started right after July~10, 2016,
and that the flux density at all frequencies was still increasing exponentially
until December~2016. We measure a time lag of $\sim$13~months between the
beginning of the IR outburst and that of the radio burst.

At the last epoch we detect a barely resolved elongation to the NE in the
direction of the jet axis, which we interpret as an expanding lobe of the
ionized gas powered by the outburst.

In order to reproduce the observed flux variations, we adopted the model
developed by Reynolds~(\cite{reyn})  under simplifying assumptions. In particular,
we assume that the jet is initially accelerated up to a distance $r_0$ where
it becomes ionized. We infer $r_0\simeq50$~au. Beyond this distance the jet expands at constant speed. We
obtain the best fit to the observed spectra at the last five epochs by varying
only two parameters: the opening angle of the jet and the ratio between
the ionized mass loss rate and the expansion speed.

Our best fits indicate that both parameters increase with time during the
burst, while the mass flux of the ionized gas through the jet remains
roughly constant. This suggests that the increase in the ionized mass loss
rate could be due to widening of the jet opening angle.

Finally, by comparison between the accretion rate computed from the IR
outburst (Caratti o Garatti et al.~\cite{cagana}) and the ionized mass loss
rate of the jet, we estimate for the first time the ionization fraction in
the jet and find a conservative upper limit of 14\%. This result strongly indicates
that the radio emission from thermal jets  very likely traces only a
negligible fraction of the jet mass, which is largely neutral.

\begin{acknowledgements}
It is a pleasure to thank Francesca Bacciotti for stimulating discussions
on jets from YSOs. We also thank the Italian ARC node and the IRAM technical
staff for their support in this project.
The research leading to these results has received funding from the European Unions
Horizon 2020 research and innovation programme under grant agreement No. 730562 [RadioNet].
This study is based on observations made under the project 16A-424
of the VLA of NRAO.
The National Radio Astronomy Observatory is a facility of the National
Science Foundation operated under cooperative agreement by Associated
Universities, Inc.
This work is also based on observations carried out under project
number E16AB with the IRAM NOEMA Interferometer. IRAM is supported
by INSU/CNRS (France), MPG (Germany), and IGN (Spain).
This paper makes use of the following ALMA data:
ADS/JAO.ALMA\#2016.A.00008.T. ALMA is a partnership of ESO (representing
its member states), NSF (USA), and NINS (Japan), together with NRC (Canada),
NSC and ASIAA (Taiwan), and KASI (Republic of Korea), in cooperation with
the Republic of Chile. The Joint ALMA Observatory is operated by ESO,
AUI/NRAO, and NAOJ.
A.C.G. and T.P.R. have received funding from the European Research Council (ERC)
under the European Union’s Horizon 2020 research and innovation programme
(grant agreement No. 743029).
\end{acknowledgements}

\appendix

\section{Turn-over frequencies of thermal jet spectra}
\label{sapp}

We want to determine the relation between the turn-over frequencies of a
typical thermal jet spectrum and the minimum ($r_0$) and maximum ($r_{\rm
max}$) radius of the jet itself. Our derivation is based on the model by
Reynolds~(\cite{reyn}).

\begin{figure}
\centering
\resizebox{8.5cm}{!}{\includegraphics[angle=0]{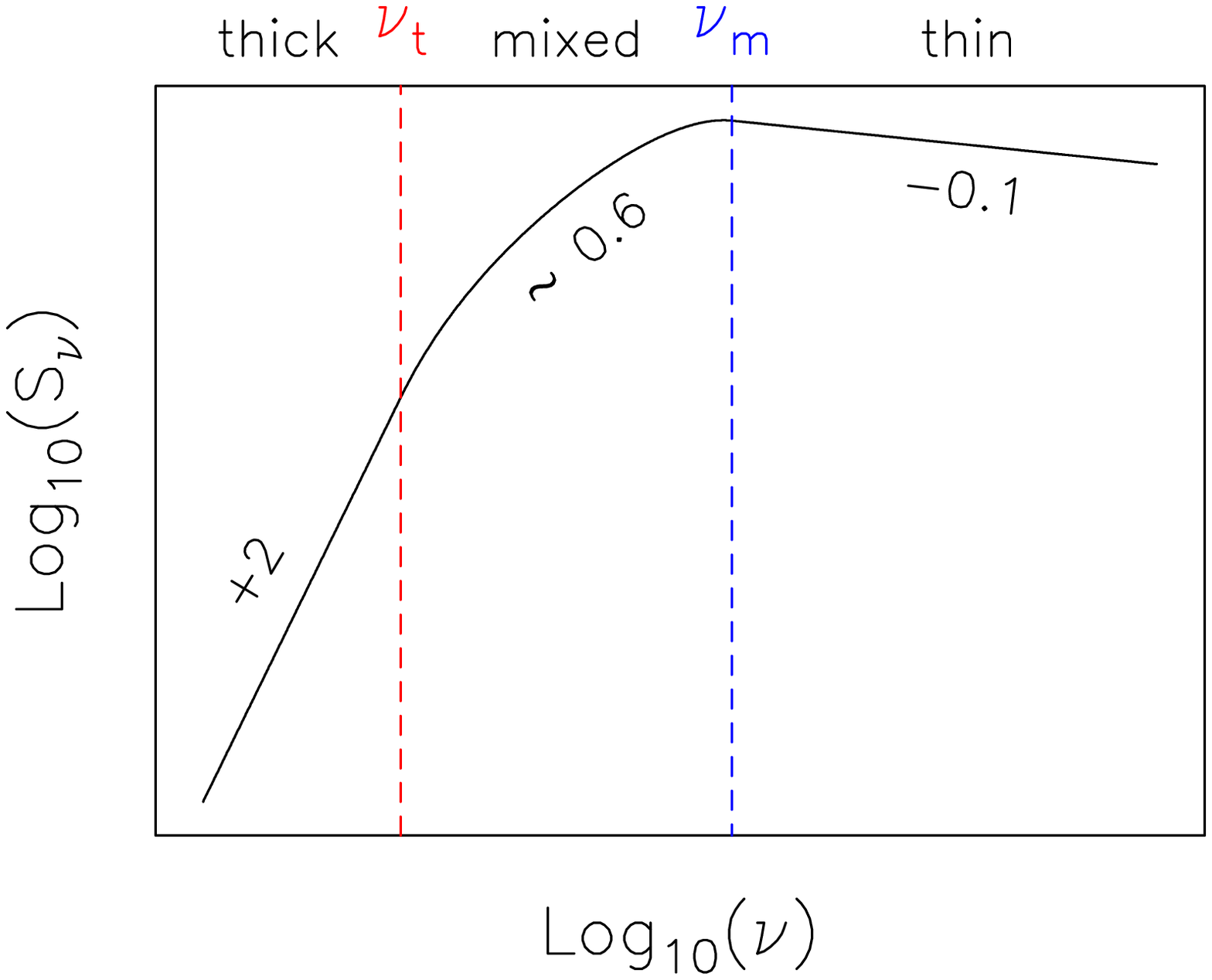}}
\caption{
Schematic radio continuum spectrum of a typical thermal jet. The two turn-over
frequencies at the border of the optically thick and thin regions of the
spectrum are indicated at the top of the figure. The regime between the
two frequencies is characterized by a mixture of thin and thick emission.
The numbers indicate the slopes in the three parts of the spectrum; 
0.6 refers to the simple case adopted by us.
}
\label{fapp}
\end{figure}

In Fig.~\ref{fapp} we show a schematic spectrum of a radio jet. We can
identify three regimes separated by the two turn-over frequencies that we
denote  $\nu_{\rm m}$ (following Reynolds' notation) and $\nu_{\rm t}$.
Since the free-free opacity decreases with frequency, there must be
a sufficiently low frequency, $\nu_{\rm t}$, below which the whole jet
(i.e. between $r_0$ and $r_{\rm max}$) is optically thick. The spectrum
in this regime is characterized by the power law $\propto\nu^2$, typical
of optically thick free-free emission.  Conversely, for a sufficiently
high frequency above $\nu_{\rm m}$, the whole jet becomes optically thin
and the power law becomes $\propto\nu^{-0.1}$.  Between $\nu_{\rm t}$ and
$\nu_{\rm m}$ the flux density is contributed by both optically thick and
thin emission, and Reynolds demonstrates that the slope takes the value of
$\sim$0.6 in the simple case of a conical, isothermal jet with density scaling
$\propto r^{-2}$ (see his Table~1).

The value of $\nu_{\rm m}$ is obtained by Reynolds in his Eq.~(13)
from the condition $r_1=r_0$, where $r_1$ is the radius at which the
free-free opacity equals 1. This assumes that the optical depth decreases
with increasing $r$, as expected if the density of the ionized gas also
decreases.

Taking into account Reynolds' expression for $\dot{M}$,
we can rewrite Reynolds' Eq.~(13) in a more convenient
form for our purposes,
\begin{equation}
 \nu_{\rm m} = \left[ \frac{2\,a_\kappa}{(\pi \mu)^2} \, (\theta_0\,r_0)^{-3} \,
 \left(\frac{x_0 \dot{M}}{\varv_0}\right)^2
 T_0^{-1.35} \, (\sin\,i)^{-1} \right]^{\frac{1}{2.1}}
 \label{enum}
,\end{equation}
where the symbols are the same as in Sect.~\ref{sdis}.

The value of $\nu_{\rm t}$ can be computed in a similar way,
from the condition $r_1=r_{\rm max}$. Taking into account that $y=r\,\sin\,i$
and using Reynolds' Eq.~(12), this condition can be written as
\begin{equation}
 \left[ \frac{2\,a_\kappa}{(\pi \mu)^2} \, (\theta_0 \,r_0)^{-3} \,
 \left(\frac{x_0 \dot{M}}{\varv_0}\right)^2
 T_0^{-1.35} \, (\sin\,i)^{-1} \, r_0^{-q_\tau} \, \nu_t^{-2.1} \right]^{-\frac{1}{q_\tau}}
 = r_{\rm max}
,\end{equation}
where $q_\tau$ is given by Reynolds' Eq.~(6). From this, we have
\begin{equation}
 \nu_{\rm t} = \left[ \frac{2\,a_\kappa}{(\pi \mu)^2} \, (\theta_0 \,r_0)^{-3} \,
 \left(\frac{x_0 \dot{M}}{\varv_0}\right)^2
 T_0^{-1.35} \, (\sin\,i)^{-1} \, \left(\frac{r_0}{r_{\rm max}}\right)^{-q_\tau} \right]^{\frac{1}{2.1}}.
 \label{enut}
\end{equation}

Finally, from Eq.~(\ref{enum}) and~(\ref{enut}) we obtain
\begin{equation}
 \frac{\nu_{\rm t}}{\nu_{\rm m}} = \left(\frac{r_{\rm max}}{r_0}\right)^\frac{q_\tau}{2.1}
\end{equation}

In conclusion, for a given value of $q_\tau$, the ratio between the two
turn-over frequencies depends only on the ratio between the outer and
inner radius of the jet, and vice versa. The case considered in our study
corresponds to $q_\tau=-3$ (see Reynolds' Table~1), hence
\begin{equation}
 \frac{r_0}{r_{\rm max}} = \left( \frac{\nu_{\rm t}}{\nu_{\rm m}} \right)^{0.7}.
    \label{eqap}
\end{equation}

\end{document}